**Title:**

**Feasibility of PET-enabled dual-energy CT imaging: First physical phantom and initial patient study results**

**Authors:**

Yansong Zhu[1], Siqi Li[1], Zhaoheng Xie[2], Edwin K. Leung[1,2,3], Reimund Bayerlein[1,2], Negar Omidvari[2], Yasser G. Abdelhafez[1], Simon R. Cherry[1,2], Jinyi Qi[2], Ramsey D. Badawi[1,2], Benjamin A. Spencer[1,2], Guobao Wang[1]

**Affiliations:**

1. Department of Radiology, UC Davis Health, Sacramento, CA 95817, U.S.A.

2. Department of Biomedical Engineering, University of California at Davis, Davis, CA 95616, U.S.A.

3. UIH America, Inc., Houston, TX 77054, U.S.A.

**Corresponding author**:

Yansong Zhu, Email: yszhu@ucdavis.edu.

**ABSTRACT**

**Purpose:** Dual-energy (DE) CT enables material decomposition by using two different x-ray energies and may be combined with PET for improved multimodality imaging. However, this increases radiation dose and may require a hardware upgrade due to the added second x-ray CT scan. The recently proposed PET-enabled DECT method allows dual-energy imaging using a conventional PET/CT scanner without the need to change scanner hardware or increase radiation exposure. Here we demonstrate the first-time physical phantom and patient data evaluation of this method.

**Methods**: The PET-enabled DECT method reconstructs a gamma-ray CT (gCT) image at 511 keV from the time-of-flight PET data with the maximum-likelihood attenuation and activity (MLAA) approach and then combines this image with the low-energy x-ray CT images to form a dual-energy image pair for material decomposition. To improve the image quality of gCT, a kernel MLAA method was developed using the x-ray CT as *a priori* information. Here we developed a general open-source implementation for gCT reconstruction and used this implementation for the first real data validation using both physical phantom study and human-subject study. Results from PET-enabled DECT were compared using x-ray DECT as the reference. Further, we applied the PET-enabled DECT method in another patient study to evaluate bone lesions.

**Results:** Compared to the standard MLAA, results from the kernel MLAA showed significantly improved image quality. PET-enabled DECT with the kernel MLAA was able to generate fractional images that were comparable to the x-ray DECT, with high correlation coefficients for both the phantom study and human subject study (R>0.99). The application study also indicates that PET-enabled DECT has potential to characterize bone lesions.

**Conclusion**: Results from this study have demonstrated the feasibility of this PET-enabled method for CT imaging and material decomposition. PET-enabled DECT shows promise to provide comparable results to x-ray DECT.

**INTRODUCTION**

Dual-energy computed tomography (DECT) performs two sets of attenuation measurements using two different x-ray energy spectra to provide quantitative information on tissue compositions [1]. This information, when combined with positron emission tomography (PET), has many potential applications. For example, it has the potential to improve attenuation correction for contrast-enhanced PET/CT imaging where the high-Z contrast media may cause inaccuracy and artifacts in PET attenuation correction if single-energy CT is used [2,3]. The method can also correct the bone fraction effect in a "bone marrow" region to improve metabolic PET quantification for bone marrow imaging [4]. Similarly, the method has the potential to measure accurate volumes of bone marrow and trabecular bone to improve PET-based predictive bone marrow dosimetry in targeted radionuclide therapy [5].

However, performing DECT imaging with an existing PET/CT system is not trivial. Direct replacement of single-energy CT with new DECT is costly because the latter requires a different scanner configuration, and its price is much higher than a single-energy CT scanner. Utilization of existing single-energy CT is possible but requires changes on data acquisition protocol and is associated with an increased radiation dose and scanning cost. As a result of these barriers, x-ray DECT imaging has not been widely explored in PET/CT molecular imaging, despite some ongoing vendor's efforts in implementing x-ray DECT on PET/CT scanners.

The PET-enabled DECT method is a recently proposed technique [6] that allows for the simultaneous PET and DECT imaging using a conventional PET/CT scanner without the need to change the scanner hardware or increase radiation exposure to patients. In this PET-enabled DECT method, a high-energy γ-ray CT (gCT) at 511 keV is obtained from time-of-flight (TOF) PET emission data. The 511-keV gCT image is then combined with the existing lower-energy x-ray CT image to form a DECT image pair for material decomposition. The reconstruction of the gCT image from PET data can be achieved using joint estimation of attenuation and radiotracer activity, for example, with the maximum-likelihood attenuation and activity (MLAA) reconstruction algorithm [7]. The MLAA topic has garnered broad research interest [7–11], but the previous attention was mainly focused on the task of attenuation correction [10,12–15]. Note that the standard MLAA reconstruction can be noisy due to the limited count level of the PET emission data. For our DECT applications, the noise from the gCT image could propagate into material decomposition, resulting in noise and artifacts in tissue fractional images. To alleviate these issues, the kernel MLAA reconstruction was proposed, in which the x-ray CT image is used as *a priori* knowledge and incorporated in the forward model of the inverse problem to guide the MLAA reconstruction [6,16]. The gCT image from kernel MLAA has shown substantially improved image quality, particularly for low-count PET emission data.

While the concept of PET-enabled DECT was proposed, the evaluation of this technique in previous work was mainly limited to 2D simulation studies [6,16]. In this work, we developed a general, open-source implementation of MLAA and kernel MLAA that allows 3D gCT imaging for clinical PET/CT scanners. With this implementation, we demonstrated the feasibility of PET-enabled DECT using one physical phantom study and two patient studies with data collected from the uEXPLORER total-body PET/CT system [17]. The results from this work represent the first real data validation of the PET-enabled DECT method.

**METHODS**

**PET-enabled DECT**

*Reconstruction of gCT image.* In TOF PET, the measured projection data $y$ is modeled using the Poisson distribution with the log-likelihood function as:

$$L(y|\lambda,\mu) = \sum_{i=1}^{N_d} \sum_{m=1}^{N_t} y_{i,m} \log \bar{y}_{i,m}(\lambda,\mu) - \bar{y}_{i,m}(\lambda,\mu), \qquad (1)$$

where $\lambda$, $\mu$ are the radiotracer activity and the gCT images to be reconstructed, respectively. $i$ and $m$ are indices for line of response (LOR) and TOF bin, respectively. $N_d$ is the total number of LORs and $N_t$ is the number of TOF bins. $\bar{y}_{i,m}(\lambda,\mu)$ is the expected projection data and can be written as:

$$\bar{y}_m(\lambda,\mu) = \text{diag}\{n_m(\mu)\}G_m\lambda + r_m, \qquad (2)$$

where $G_m$ is the PET detection probability matrix. $r_m$ is the expected contribution from random and scatter events. $n_m$ is the normalization factor and is given as:

$$n_{i,m} = c_{i,m} \cdot \exp(-[A\mu]_i), \qquad (3)$$

where $c_{i,m}$ denotes a multiplicative factor excluding the attenuation correction factor. $A$ is the system matrix for the corresponding transmission scan. The standard MLAA algorithm seeks to solve the maximum-likelihood estimation problem:

$$\hat{\lambda}, \hat{\mu} = \arg \max_{\lambda \geq 0, \mu \geq 0} L(y|\lambda, \mu), \qquad (4)$$

which tends to result in noisy gCT images [11–13]. To address this issue, the kernel MLAA was further developed [6].

In kernel MLAA, the gCT image $\mu$ is represented using a kernel representation [6]:

$$\mu = K\alpha, \qquad (5)$$

where $\alpha$ is the kernel coefficient image. $K$ is a sparse kernel matrix that is built by using the x-ray CT as *a priori* information. In this work, a radial Gaussian kernel is used to build the kernel matrix with the $(i,j)$th element of $K$ being

$$\kappa(f_i, f_j) = \exp\left(-\frac{\|f_i - f_j\|^2}{2\sigma^2}\right), \qquad (6)$$

where $f_i$, and $f_j$ are $3 \times 3 \times 3$ image patches centered at the $i$, $j$th voxel, respectively. $\sigma$ is a hyper-parameter and is fixed as 1 in this work. The kernel matrix was constructed by selecting 50 neighbors for each voxel using K-Nearest-Neighbor search. With this kernel representation, the optimization problem in equation (4) becomes the following kernel MLAA reconstruction problem:

$$\hat{\lambda}, \hat{\alpha} = \arg \max_{\lambda \geq 0, \alpha \geq 0} L(y|\lambda, K\alpha). \qquad (7)$$

The optimization problem in equation (7) can be solved by updating $\lambda$ and $\alpha$ alternatively as:

$$\hat{\lambda} = \arg \max_{\lambda \geq 0} L(y|\lambda, K\hat{\alpha}), \qquad (8)$$

$$\hat{\alpha} = \arg \max_{\alpha \geq 0} L(y|\hat{\lambda}, K\alpha). \qquad (9)$$

The $\lambda$-step can be solved with the standard expectation-maximum (EM) algorithm, and the $\alpha$-step can be solved using an optimization transfer method with sparable quadratic surrogate [6,18]. Once $\hat{\alpha}$ is estimated, the final gCT image is given by $\hat{\mu} = K\hat{\alpha}$.

*Open-source implementation.* We implemented the standard MLAA and kernel MLAA reconstruction algorithms based on an open-source reconstruction platform: Customizable and Advanced Software for Tomographic Reconstruction (CASToR) package [19]. The projector code of CASToR was extracted and compiled as a mex file. Our MATLAB-based in-house implementation of standard MLAA and kernel MLAA codes could then use this generic projector to reconstruct 3D gCT images of various PET/CT systems. When running the code for a new scanner, the scanner geometry file can be added following the description in CASToR's documentation. A MATLAB function is also used to create a table for different scanners, where other data description parameters (for example, data type and TOF-related parameters) required by the CASToR projector can be provided.

*Multi-material decomposition with gCT and x-ray CT.* With gCT $\boldsymbol{\mu}$ and x-ray CT $\boldsymbol{x}$, tissues can be decomposed into a set of basis materials, such as air (A), bone (B), and water (W). This is described as:

$$\boldsymbol{u}_j \triangleq \begin{pmatrix} x_j \\ \mu_j \end{pmatrix} = \boldsymbol{U}\boldsymbol{\rho}_j, \qquad (10)$$

where $j$ is voxel index. $\boldsymbol{U} = \begin{pmatrix} x_A & x_B & x_W \\ \mu_A & \mu_B & \mu_W \end{pmatrix}$ denotes attenuation coefficients for each basis material measured at the low and high energies. The coefficients vector $\boldsymbol{\rho}_j = (\rho_{j,A} \ \rho_{j,B} \ \rho_{j,W})^T$ gives tissue fractions for each basis material at voxel $j$ and subject to $\sum_k \rho_{j,k} = 1$, where $k = \{A, B, W\}$. The tissue fraction coefficients can be obtained by solving the least-square optimization problem:

$$\widehat{\boldsymbol{\rho}}_j = \arg\min_{\boldsymbol{\rho}_j \geq 0} \|\boldsymbol{u}_j - \boldsymbol{U}\boldsymbol{\rho}_j\|^2. \qquad (11)$$

By filling the estimated tissue fraction coefficients $\widehat{\boldsymbol{\rho}}_j$ in each voxel, we get the fractional images corresponding to the basis materials, including air, bone and water.

**Validation using phantom scans and X-ray DECT.**

*Data acquisition on uEXPLORER.* We demonstrated the PET-enabled DECT method first using a physical phantom scan. A cylindrical phantom of radius 14 cm with five compartments was used in this study (Fig. 1). The phantom was filled with water in the background, and the four inserts were filled with (1) Styrofoam balls as lung tissue-equivalent material, (2) distilled water, and (3-4) salt water, which contains about 0.31 kg salt in each of the two compartments, as shown in Fig. 1(c). All five compartments were uniformly filled with $^{18}$F-FDG solutions, with total activity at about 185 MBq, as shown in Fig. 1(b). Due to the filling of Styrofoam balls, the filling volume of FDG solution in insert 1 is about 1/3 compared to inserts 2-4, which leads to the lower activity in this compartment. Additional attenuation materials composed of fat tissue-equivalent material (the blue material in Fig.1(a)) and bovine rib bones were wrapped around the phantom. The phantom was scanned with the uEXPLORER PET/CT system at UC Davis which has a TOF resolution of 505 ps [17]. Two minutes of TOF PET data from one uEXPLORER unit (24 cm axial field of view (AFOV)) were used for image reconstruction. This was based on the consideration of computational efficiency for the very first validation and the typical AFOV (15-30 cm) of a conventional clinical PET scanner. In addition, two x-ray CT scans, at 140 kVp (180 mAs) and 80 kVp (245 mAs) respectively, were obtained to perform x-ray DECT and provide the reference for material decomposition. The CT images were down sampled to 4×4×4 mm$^3$ to match with the voxel size from PET-enabled DECT.

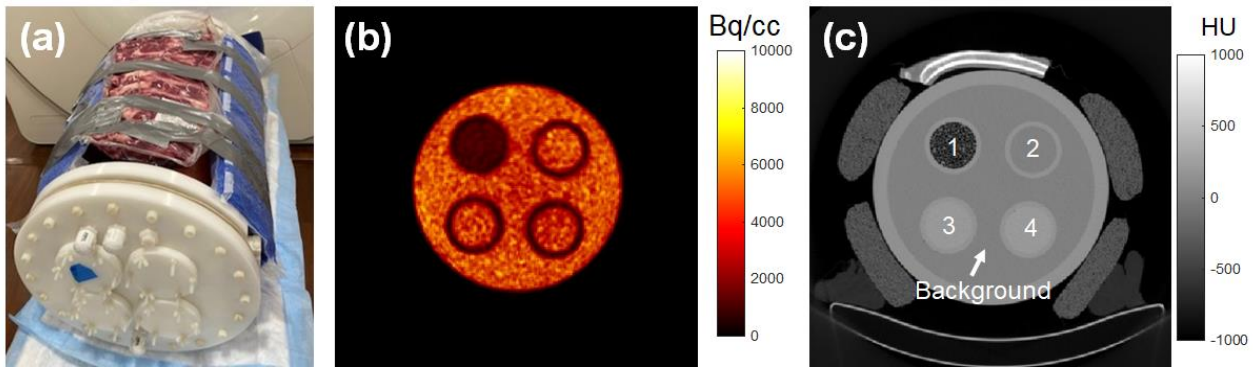

**Fig. 1** Illustration of the phantom used for the phantom study. (a) Picture of the physical phantom, and the transverse slices of (b) the activity image and (c) the 140 kVp CT image.

*Image reconstruction and data correction*. For gCT image reconstruction, both the standard MLAA and the kernel MLAA were applied. Since the 80 kVp CT image has higher contrast between different tissues [20], we used the 80 kVp CT image to generate the kernel matrix for the kernel MLAA, and to perform dual-energy imaging together with the reconstructed gCT images. In addition, a bilinear scaling method was used to convert the x-ray CT image to 511 keV attenuation image [21], which was used to initialize both the standard and kernel MLAA algorithms to accelerate convergence, as suggested by [6].

For data correction, we extracted the normalization factors and the estimated scatter and random events from the vendor's software. As shown in previous studies [22–25], the quantitative accuracy of MLAA reconstruction depends on the quality of multiple data correction factors. The same effect is also expected for kernel MLAA. To reduce the potential errors from scatter correction, the scatter events obtained from the vendor's software were further scaled by a global scale factor that was obtained by a maximum-likelihood joint estimation of scatter scaling [26]. The gCT image was then reconstructed with an image size of 150×150×60 and a voxel size of 4×4×4 mm³. Twenty iterations with 20 subsets were used to reconstruct the gCT images for both standard MLAA and kernel MLAA methods.

*Evaluation metric.* Quantitative evaluation was performed for multi-material decomposition (MMD) results of PET-enabled DECT with the standard MLAA and the kernel MLAA. Region of interests (ROIs) were placed in the water and salt-water regions for the water fractional images, and in the bone region for the bone fractional images. By changing the number of iterations, we plotted the ROI mean vs. background noise curves. The voxel-wise noise was computed in a uniform region as:

$$noise = \frac{1}{c^{ref}} \sqrt{\frac{\sum_{i=1}^{N}(c_i - \bar{c})^2}{N-1}}, \quad (12)$$

where $c^{ref}$ is the average ROI intensity extracted from the reference image, which is the fractional image generated from x-ray DECT in this study. $\bar{c} = \frac{1}{N} \sum_{i=1}^{N} c_i$ is the average ROI intensity, and $N$ is the total number of voxels in the ROI.

We then compared the quantitative results from PET-enabled DECT and the reference x-ray DECT. We computed the ROI means in different regions (including the water, salt-water, bone, lung-equivalent material, and fat-equivalent material regions) in the water and bone fractional images from these two DECT methods. Supplemental Fig. 1 showed the ROI placement for different regions in the phantom study. Bland-Altman plots were used to test the agreement between the two DECT methods. The scatter plots were also plotted and Pearson correlation coefficient was computed between ROI measures from the two methods.

**Validation using patient scans and X-ray DECT.**

*Data acquisition and image reconstruction.* We further validated PET-enabled DECT with a real patient scan. In this validation study, data from a single patient with metastatic genitourinary cancer patient was utilized in this study. Prior Ethics Committee and Institution Review Board approval and informed consent were obtained (IRB#: 1374902). The patient was injected with ~370 MBq of [18]F-FDG. PET emission scan data from 55-60 min post-injection was used for reconstruction of the gCT images. Apart from the PET emission data, two x-ray CT scans were collected sequentially before the injection of FDG, one scan at 140 kVp (50 mAs), and the other at 80 kVp (100 mAs). Because the 80 kVp CT image was relatively noisy, it was first denoised using a deep-learning-based algorithm [27] to improve the image quality. The denoised 80 kVp x-ray CT image was registered to the 140 kVp x-ray CT image to avoid motion artifact in the x-ray DECT. For PET-enabled DECT, the same data correction process was applied for the patient data. Similar to the phantom study, the 80 kVp CT image was applied to guide gCT reconstruction and to perform dual-energy imaging together with the gCT image.

*Evaluation metric.* Similar to the phantom study, we first performed quantitative evaluation for MMD results of PET-enabled DECT with the standard MLAA and the kernel MLAA. ROIs were placed in the liver region for the water fractional images, and in the humerus bone for the bone fractional images. ROI mean vs. background noise curves were plotted for the two PET-enabled DECT methods. The quantitative results from PET-enabled DECT and the reference x-ray DECT were also compared by placing ROIs in different tissues, as shown in supplemental Fig. 2. ROI quantification was performed by averaging the fractional values over all ROIs per tissue. The agreement between PET-enabled DECT and x-ray DECT was validated using Bland-Altman plot and scatter plot.

**Application to patient scans with bone lesions.**

*Data acquisition and image reconstruction.* We applied the PET-enabled DECT method on another metastatic genitourinary cancer patient who had three bone metastatic lesions to show the potential application of PET-enabled DECT to characterize bone lesions. We collected the FDG-PET data together with a 140 kVp CT image in this study. Data acquisition process for this patient was similar to the aforementioned patient study but no x-ray DECT scan was performed. The gCT image was reconstructed with the kernel MLAA method, and fractional images were generated using the gCT and 140 kVp CT image pair.

*Evaluation metric.* ROIs were placed in the three bone lesions. In addition, non-lesion ROIs were placed in the normal tissue at the contralateral side of the lesion and at normal nearby bony tissue that is close to the lesions. We placed total 9 ROIs to cover the three bone lesions and 10 ROIs in non-lesion tissue. All the ROIs are in the same size of $10\times10\times10$ mm$^3$. Supplemental Fig. 3 shows an example of ROI placement in lesion and non-lesion regions for one of the bone lesions. Average bone fractional values were calculated for these ROIs. Boxplot was used to show the group differences between lesion and non-lesion groups. Mann–Whitney U test was also performed between these two groups.

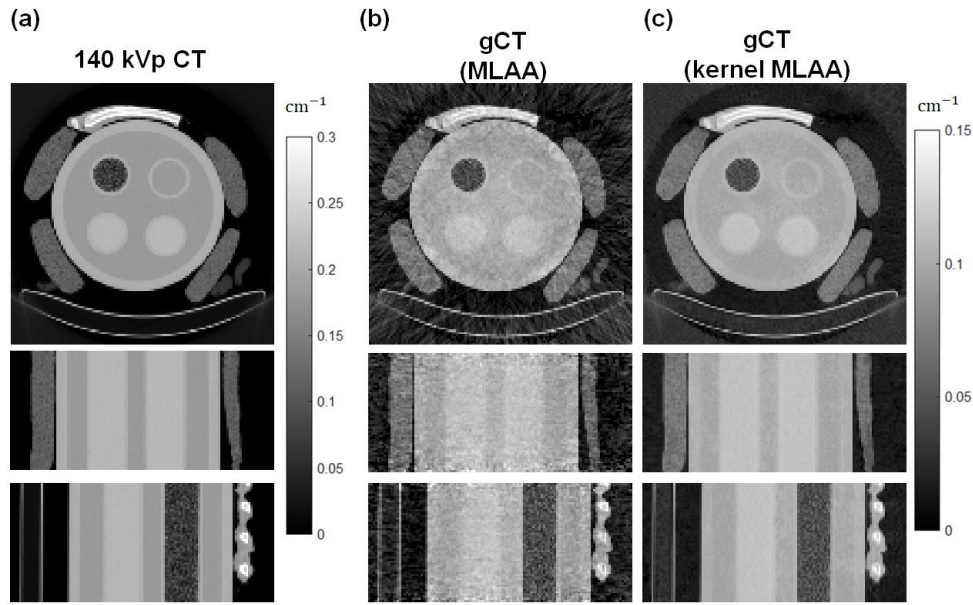

**Fig. 2** Transverse (top row), coronal (middle row), and sagittal (bottom row) views of (a) 140 kVp x-ray CT, (b) reconstructed gCT (511 keV) images with the standard MLAA, and (c) kernel MLAA method for the phantom study.

## RESULTS

**Validation using phantom scans.**

Fig. 2 shows the x-ray CT image and the reconstructed gCT images using the standard MLAA and kernel MLAA methods. The gCT image from the standard MLAA showed substantial image noise artifacts, which influences the visualization of structural details in the image, such as the bones. The gCT image from the kernel MLAA method significantly reduced the noise and revealed more prominent structural information.

With the gCT image, we then performed MMD with PET-enabled DECT and compared to the x-ray DECT method. Fig. 3 shows the water and bone fractional images for the phantom data. It can be observed that due to the propagated noise from the gCT image, the fractional images from PET-enabled DECT with the standard MLAA contain much more image noise compared to the images from the x-ray DECT. In comparison, PET-enabled DECT with the kernel MLAA generated fractional images that were visually comparable to the x-ray DECT.

Fig. 4(a) shows the curves of ROI mean vs. background noise to compare the quantitative performance of PET-enabled

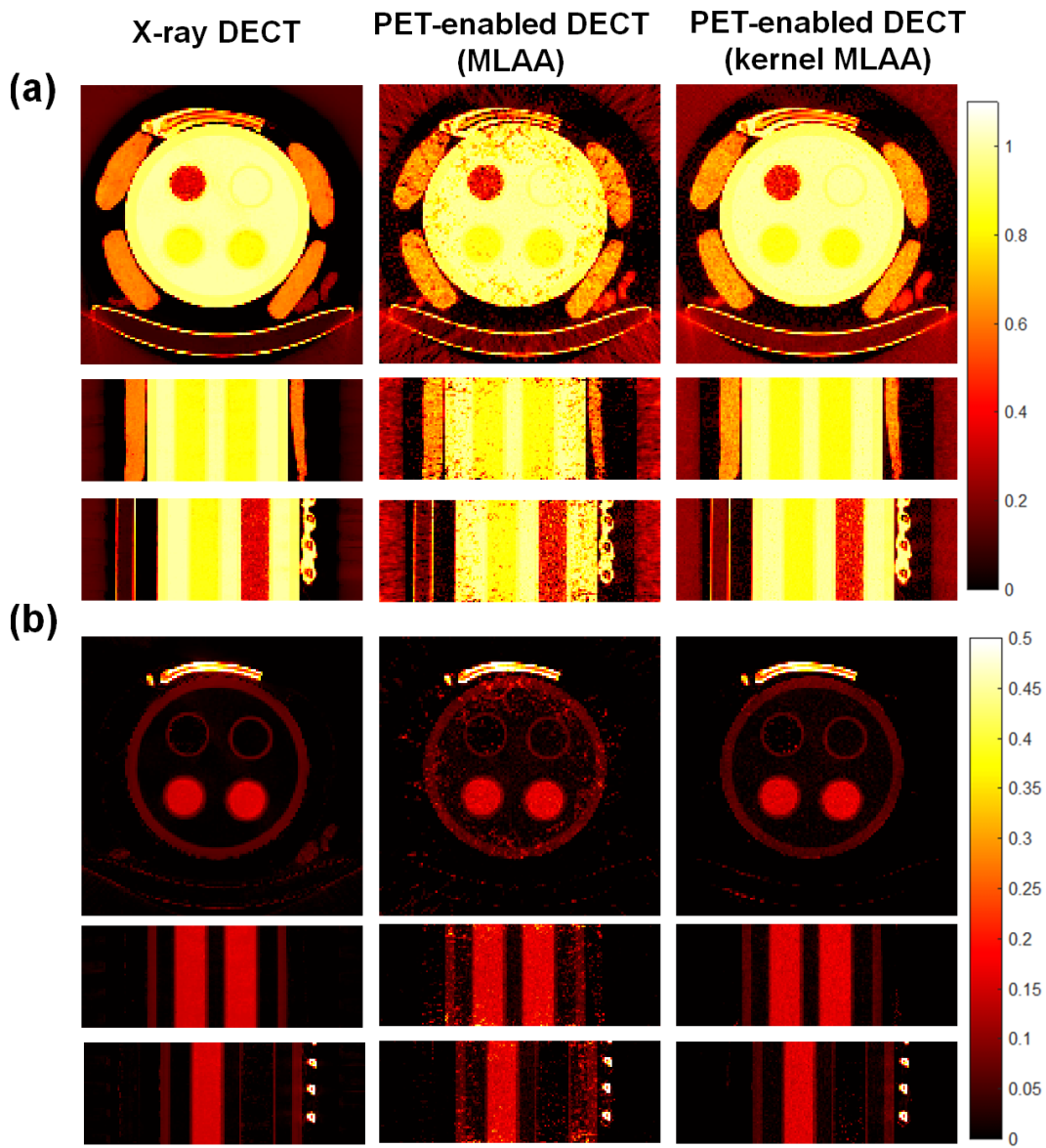

**Fig. 3** 3D views of (a) water fractional images, and (b) bone fractional images generated by x-ray DECT (left column), PET-enabled DECT with the standard MLAA (middle column), and with kernel MLAA (right column) for the phantom study.

DECT with the standard MLAA and kernel MLAA. As the iteration number increases, the standard MLAA led to a decreased average water fractions for water and salt-water ROIs in the water fractional images, and increased bone values in the bone fractional images. The noise in the fractional images from the standard MLAA substantially increased as the number of iterations increases, which is consistent with the visual observations from Fig. 3. In comparison, the kernel MLAA kept the noise at a lower level.

Fig. 4(b) shows the Bland-Altman plots for ROI quantification between PET-enabled DECT (with the kernel MLAA) and x-ray DECT. The scatter plots were shown in Supplemental Fig. 4. Supplemental Fig. 5 shows the difference images between PET-enabled DECT and x-ray DECT for a voxel-wise comparison. In general, we observed a good quantitative agreement between PET-enabled DECT and x-ray DECT in both the water fractional image and the bone fractional image. Relative larger differences were observed in low-count regions such as in the lung-equivalent material and fat-equivalent material.

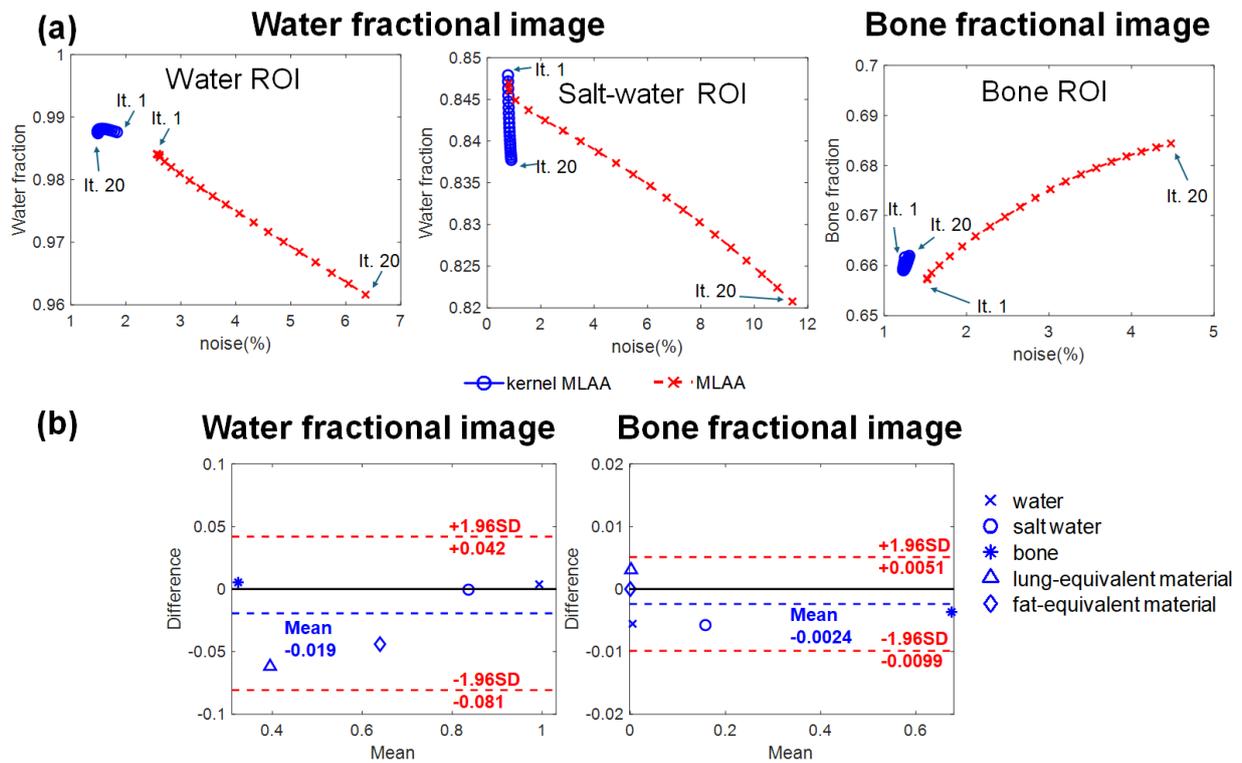

**Fig. 4** Quantitative evaluation results of the phantom study. (a) Comparison between PET-enabled DECT with MLAA and kernel MLAA methods using the ROI mean vs. background noise curves in the water and bone fractional images for different ROIs. The directions of the iteration numbers are also indicated in the curve ("It." stands for "iteration"). (b) Bland-Altman plots for comparison between x-ray DECT and PET-enabled DECT (with kernel MLAA) in the water and bone fractional images. The dashed red lines show the range of 1.96 standard deviations that represent the range of 95% confidence interval.

**Validation using patient scans.**

Fig. 5(a) shows the reconstructed gCT images together with the 140 kVp x-ray CT image. Consistent with the observation from the phantom study, the gCT image from the standard MLAA was nosier compared to the x-ray CT image and gCT image from kernel MLAA. Kernel MLAA substantially reduced the image noise and generated the gCT images with improved visual image quality.

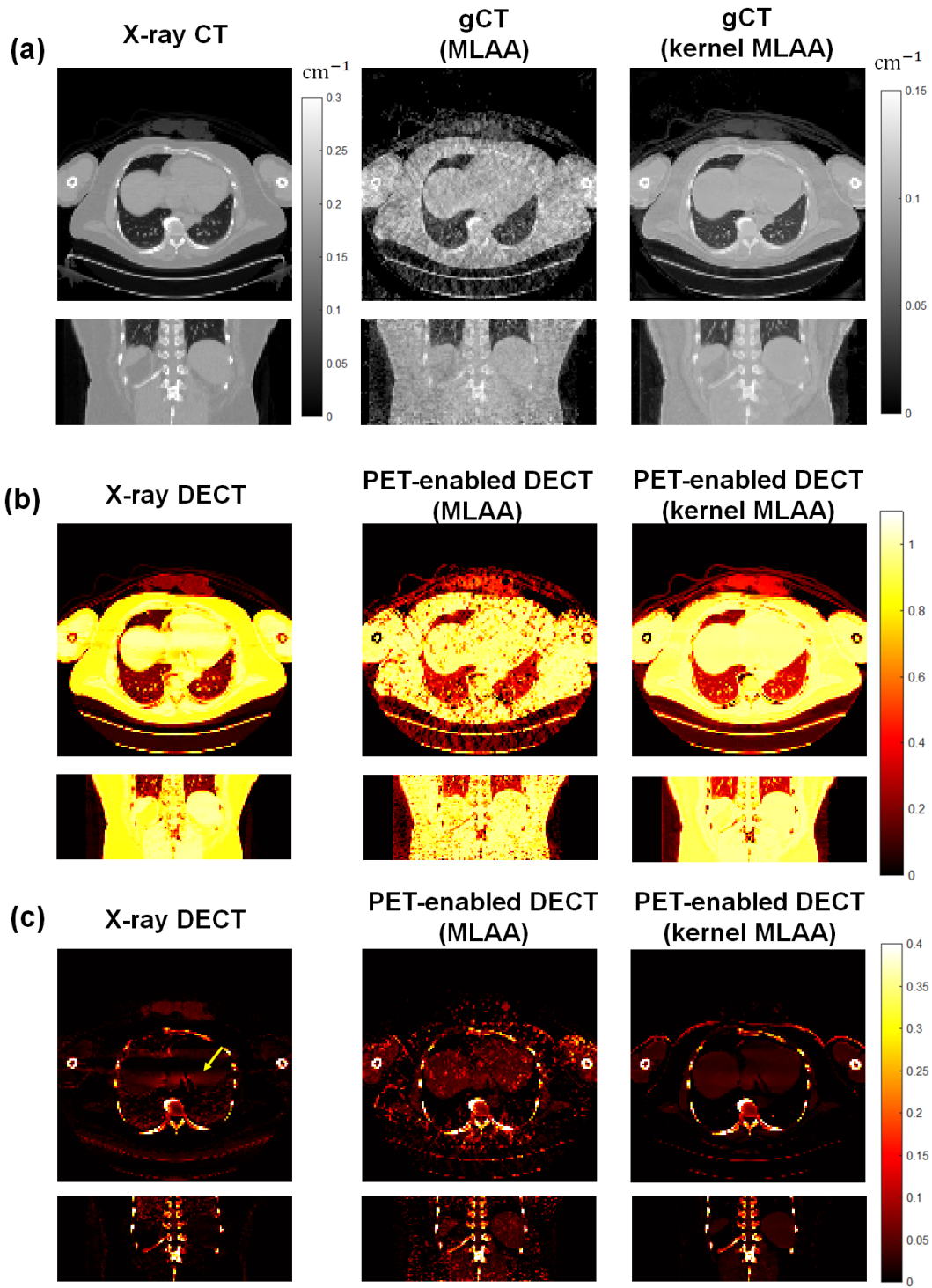

**Fig. 5** Transverse and coronal slices of (a) gCT and x-ray CT images, (b) water fractional images, and (c) bone fractional images obtained using x-ray CT or x-ray DECT (left column), standard MLAA (middle column) and kernel MLAA (right column). The arrow in (c) indicates the streak artifact shown in the x-ray DECT image.

The water and bone fractional images obtained with MMD are shown in Fig. 5(b) and Fig. 5(c), respectively. While the fractional images by the standard MLAA were very noisy, the images by the kernel MLAA showed a substantial quality improvement and were closer to the x-ray DECT reference. In addition, streak artifacts due to the beam hardening effect can be observed from the x-ray DECT, for example, in the bone fractional image. In comparison, the gCT image in PET-enabled DECT is obtained from single-energy photon at 511 keV. Such artifact is less apparent in the images obtained with PET-enabled DECT, either using MLAA method or kernel MLAA method.

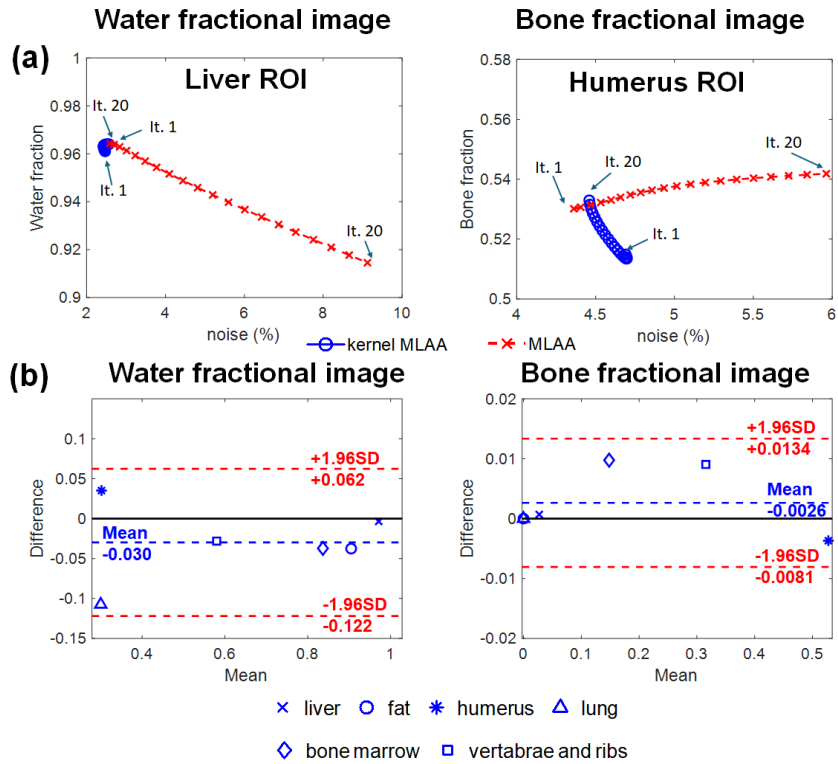

**Fig. 6** Quantitative evaluation results of the validation study with patient scan. (a) Comparison between PET-enabled DECT with MLAA and kernel MLAA methods using ROI mean vs. background noise curves. The directions of the iteration numbers are also indicated in the curve ("It." stands for "iteration"). (b) Bland-Altman plots for comparison between x-ray DECT and PET-enabled DECT (with kernel MLAA) in water and bone fractional images. The dashed red lines show the range of 1.96 standard deviations that represent the range of 95% confidence interval.

Fig. 6(a) shows the quantitative results to compare the performance of PET-enabled DECT with the standard MLAA and the kernel MLAA methods using the patient data. As the number of iterations increased, the standard MLAA showed increased noise with reduced average water fractions in the liver region. As a comparison, the kernel MLAA showed a more stable ROI quantification with a lower noise level. For the bone fractional images, both the standard MLAA and kernel MLAA showed increased ROI mean as the number of iterations increased, but the kernel MLAA achieved a lower noise level compared to MLAA.

Fig. 6(b) shows the Bland-Altman plots for ROI quantification from different tissues between x-ray DECT and PET-enabled DECT with kernel MLAA. Supplemental Fig. 6 shows the scatter plots. The difference image between PET-enabled

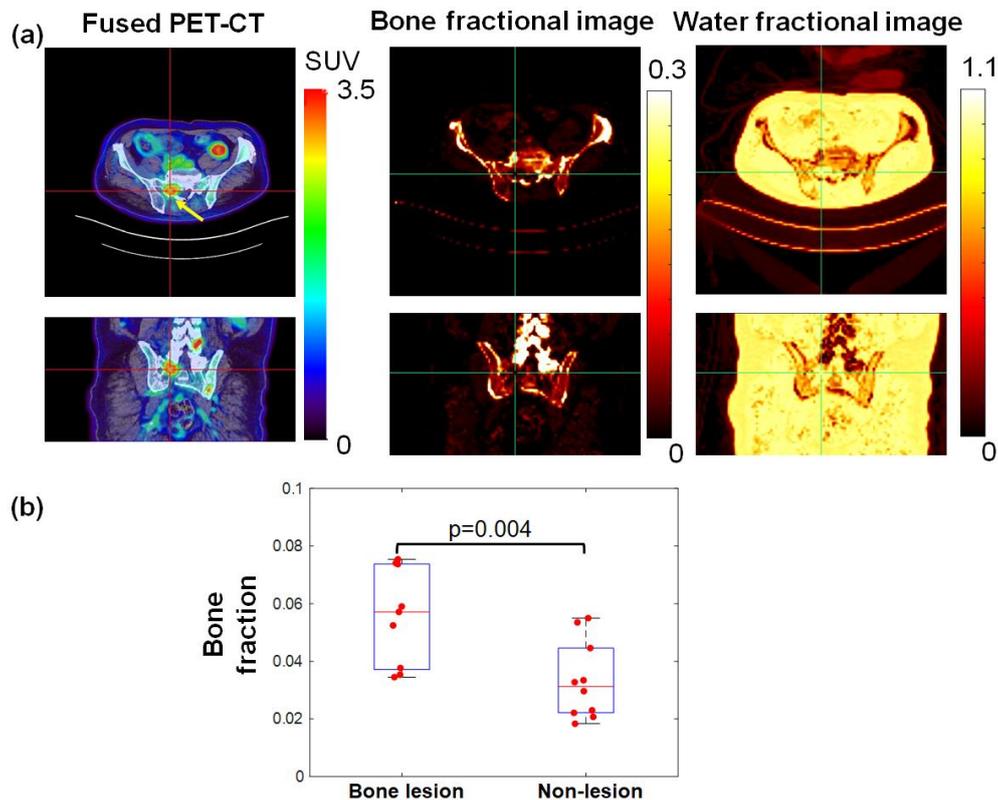

**Fig. 7** Results for the bone lesion evaluation study using patient data. (a) Transverse and coronal slices of fused PET-CT (left column), bone fractional image (middle column) and water fractional image (right column) obtained from PET-enabled DECT. (b) Boxplots of average bone fractional values for lesion and non-lesion groups. The p value of the Mann–Whitney U test was also indicated between the two groups.

DECT and x-ray DECT is shown in supplemental Fig. 7. Overall, PET-enabled DECT was quantitatively comparable to x-ray DECT. Some discrepancies were observed in lung region and in the region where x-ray DECT was affected by streak artifacts.

**Evaluation of bone lesions**

The results for evaluation of bone lesions using PET-enabled DECT are summarized in Fig. 7. Fig. 7(a) shows the fused PET-CT image for one of the bone lesions with the corresponding water and bone fractional images generated using PET-enabled DECT. Supplemental Fig. 8 shows the images for the other two lesions. It can be observed that our PET-enabled DECT successfully generated high-quality fractional images.

Fig. 7(b) shows the boxplots of quantitative bone fraction in the bone lesion and non-lesion regions. The bone lesions have higher average bone fraction values compared to the non-lesion tissues. The p value of the Mann–Whitney U test was 0.004 between the two groups. This may indicate the bone lesions show different bone fractional values compared to normal tissues.

**DISCUSSION**

In this work, we developed a 3D implementation that allows PET-enabled DECT to be performed on TOF-capable clinical PET/CT scanners. We then validated this technique with both physical phantom data and patient data that were collected from the uEXPLORER total-body PET/CT scanner. X-ray DECTs for both studies were also obtained to provide a reference.

For both the validation studies using the phantom and the patient data, PET-enabled DECT with the kernel MLAA reconstruction was able to generate images that demonstrated comparable quality to those from x-ray DECT. In comparison, the results from the standard MLAA reconstruction had substantially higher image noise, as also confirmed by the trade-off curves of ROI mean versus background noise in Fig. 4(a) and Fig. 6(a). In addition, while streak artifacts are observed from the images generated by x-ray DECT (as shown in Fig. 5), PET-enabled DECT could generate fractional images that showed less artifacts. We further applied PET-enabled DECT on a patient scan with metastatic bone lesions. The result shows that the bone fraction in the bone lesions was higher compared to the normal tissues, indicating that PET-enabled DECT has potential to provide additional useful information on top of current PET/CT.

We did observe some discrepancies between the x-ray DECT and PET-enabled DECT results. For example, in Fig. 5(b) and Fig. 6 (b), higher lung values showed in the water fractional images for the PET-enabled DECT as compared to the images from x-ray DECT for patient data. While we did not see such discrepancies in our previous simulation studies where similar voxel size and the same basis materials were used as in this study [6,16], the differences may be caused by the insufficient accuracy of data correction. Although we have applied an estimated global scaling factor to refine scatter correction, it may be still insufficient especially for the lower-count regions, such as the lungs. Plane-dependent scatter scaling will be explored to further improve quantitative accuracy in these regions [28]. In addition, it has been shown that other factors, including crystal efficiency, TOF offset, and TOF resolution could also influence the quantitative accuracy of MLAA [22,23,25]. Previous work has shown that refined calibration of these factors could improve both activity and attenuation reconstructions for MLAA [23,25]. We expect that the benefits of these corrections will be extendable to gCT imaging for PET-enabled DECT.

The patient study in this work had a small sample size, involving only two patients, including one with three bone lesions. In this very first feasibility study for PET-enabled DECT, our aim is to show that this technique can provide valuable tissue-composition information from a PET/CT scan, which has never been demonstrated in patients before. As one of the most common types of metastatic cancer, bone metastases could occur from almost all types of primary cancer, but with the highest incident rate for patients with advanced breast or prostate cancer [29]. Several studies have shown that x-ray DECT has potential to improve the diagnostic of bone lesions with virtual non-calcium map and iodine contrast [30–32]. We will conduct a further evaluation study with a large sample size in our future work to validate this technique as well as its potential applications. For the first 3D real data validation, the gCT reconstruction and PET-enabled DECT were performed with an isotropic 4x4x4 mm$^3$ voxel size in this work, which matches a commonly used voxel size used in PET research, but is coarser than a typical CT voxel size (e.g., 1 mm in each dimension). Our ongoing work includes the development of super-resolution techniques to enable high spatial resolution gCT imaging [33]. In addition, the kernel MLAA reconstruction can be further improved by exploiting the power of deep learning [34]. We will integrate these techniques together in the future.

**CONCLUSION**

We developed a general implementation of gCT imaging for PET-enabled DECT to be performed with clinical PET/CT scanners. With this, we conducted the first real data validation of this technique with physical phantom and clinical patient data. The results indicated that the kernel MLAA reconstruction could substantially improve image quality as compared to the standard MLAA approach. PET-enabled DECT shows promise to provide comparable results as x-ray DECT and to provide additional useful information on top of current PET/CT. This technique provides a practical way to incorporate tissue composition information into PET/CT without adding an additional dose and scan time, and can be important in quantitative modeling of PET data.

**Acknowledgements:**

The authors thank the EXPLORER Molecular Imaging Center team and Dr. Lorenzo Nardo for assisting in patient data acquisition, and Dr. Tony Seibert for his assistance in setting up the x-ray dual-energy CT imaging protocol.

**Funding:**

This work was supported in part by NIH R21 EB027346.

**Author contributions:**

GW and YZ conceived the concept and designed the study. YZ developed the implementations, conducted the evaluations, and analyzed the results. BAS designed and performed the phantom scans. SL, ZX, EKL, RB, NO, YGA, SRC, JQ, and RDB contributed to the study methods and materials. YGA and RDB also contributed to data interpretation. The first draft of the manuscript was written by YZ and revised by GW, and all authors commented on previous versions of the manuscript. All authors read and approved the final manuscript.

**Ethics declarations**

**Conflict of interest**

UC Davis has a revenue-sharing agreement with United Imaging Healthcare. No other potential conflicts of interest relevant to this article exist.

**Ethical approval**

All procedures performed in studies involving human participants were in accordance with the ethical standards of the institutional and/or national research committee and with the 1964 Helsinki Declaration and its later amendments or comparable ethical standards. This study was approved by Institution Review Board at University of California, Davis (IRB#: 1374902).

**Consent to participate**

Informed consent was obtained from all participants enrolled in the study.

**Consent for publication**

Consent to publish has been received from all participants.

**Supplemental Data**

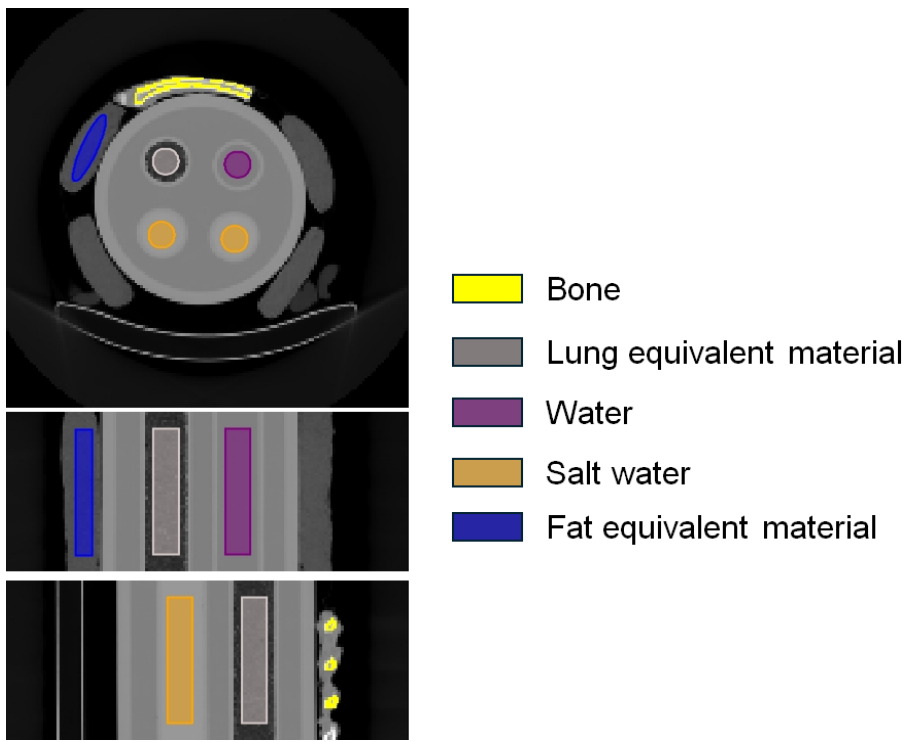

**Supplemental Fig. 1** ROI placement for quantitative evaluation of the validation study using phantom data.

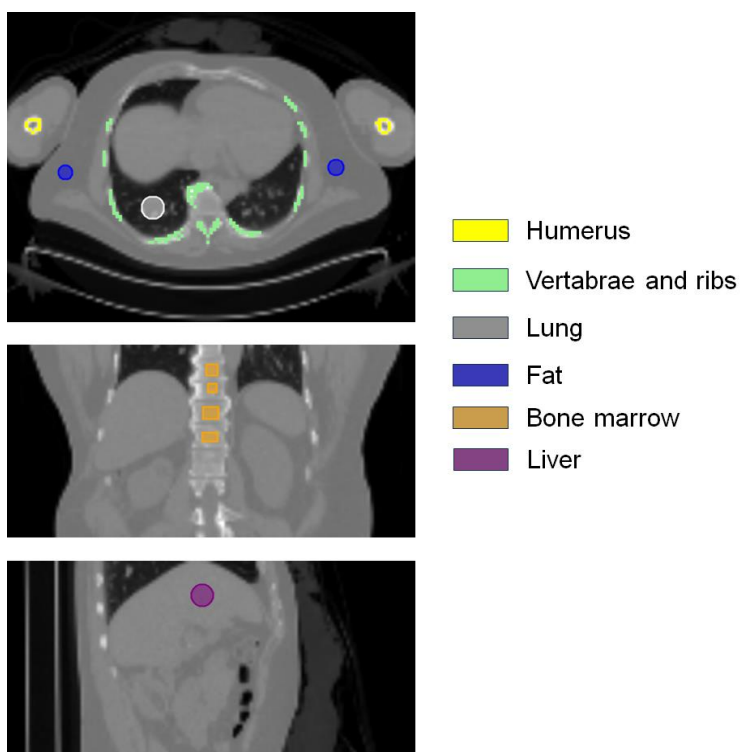

**Supplemental Fig. 2** ROI placement for quantitative evaluation of the validation study using patient data.

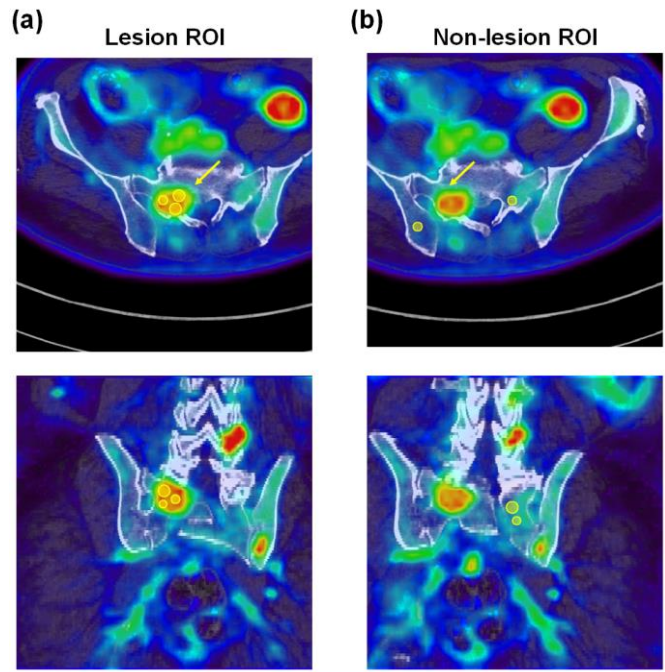

**Supplemental Fig. 3** Transverse and coronal slices for an example of (a) bone lesion ROI placement and (b) Non-lesion ROI placement. The arrow indicates the location of the bone lesion. Note that all ROIs are spherical with a diameter of 10 mm.

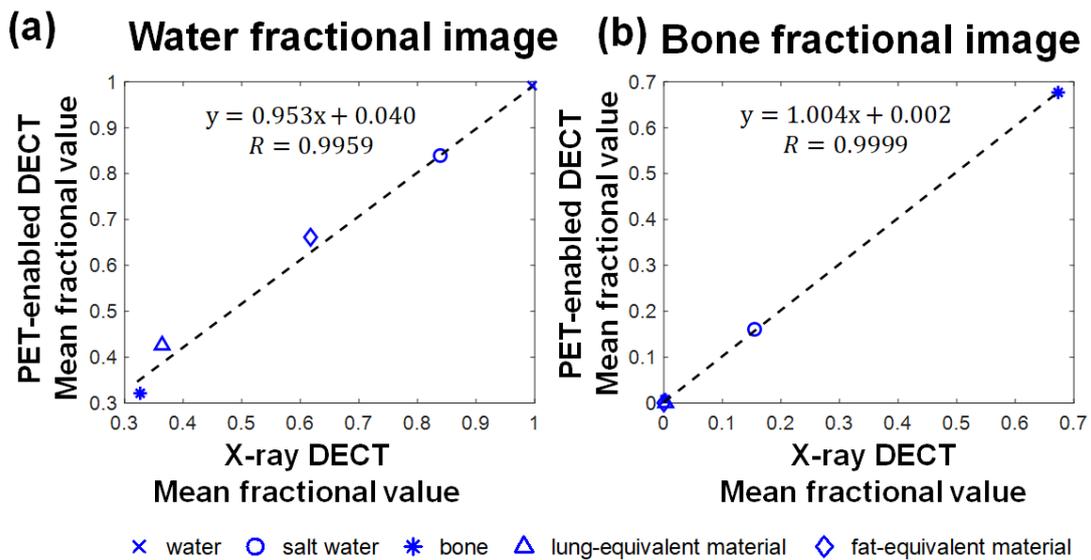

**Supplemental Fig. 4** Scatter plots for comparison between x-ray DECT and PET-enabled DECT (with kernel MLAA) with ROIs in different materials in (a) water fractional image and (b) bone fractional image for the phantom data. The dashed lines represent the linear regression fit. The fitting equation and the correlation coefficient R are also shown in the figure.

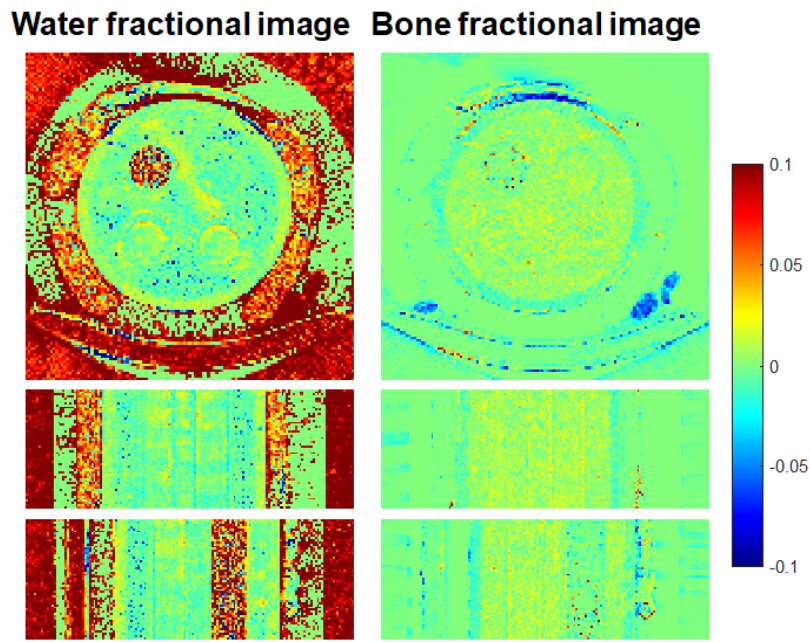

**Supplemental Fig. 5** Difference image between x-ray DECT and PET-enabled DECT (with kernel MLAA) for water fractional image (left column) and bone fractional image (right column) for the phantom study.

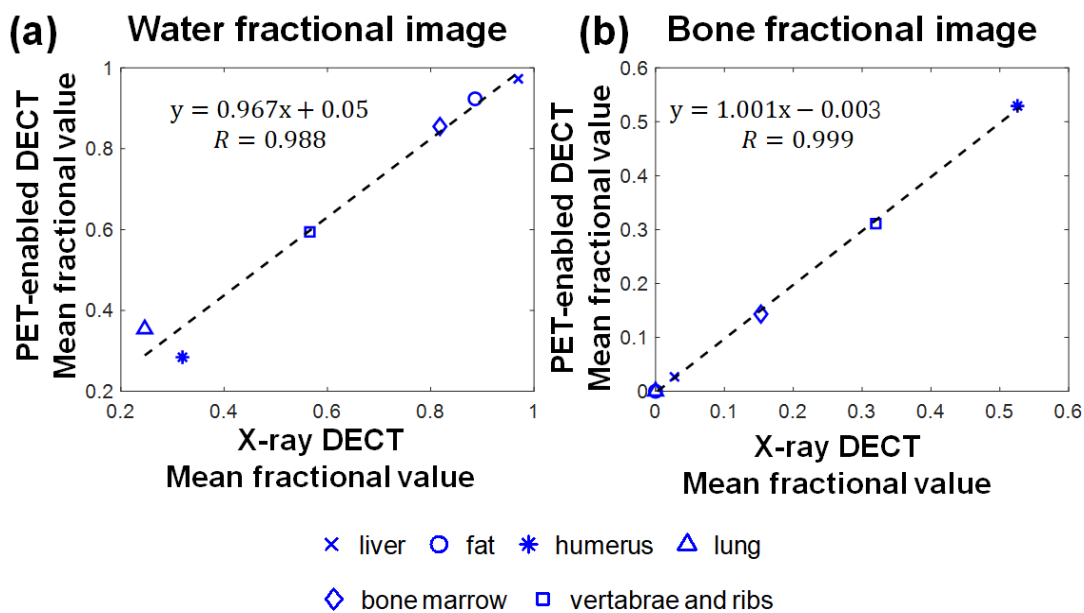

**Supplemental Fig. 6** Scatter plots for comparison between x-ray DECT and PET-enabled DECT (with kernel MLAA) with ROIs in different tissues in (a) water fractional image and (b) bone fractional image for the validation study with patient data. The dashed lines represent the linear regression fit. The fitting equation and the correlation coefficient R are also shown in the figure.

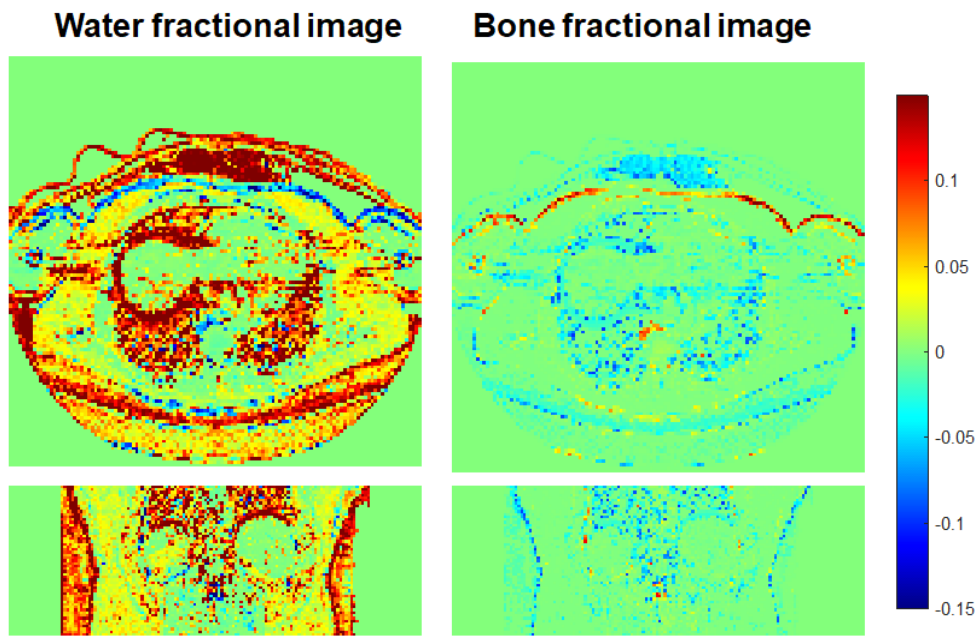

**Supplemental Fig. 7** Difference image between x-ray DECT and PET-enabled DECT (with kernel MLAA) for water fractional image (left column) and bone fractional image (right column) for the patient study.

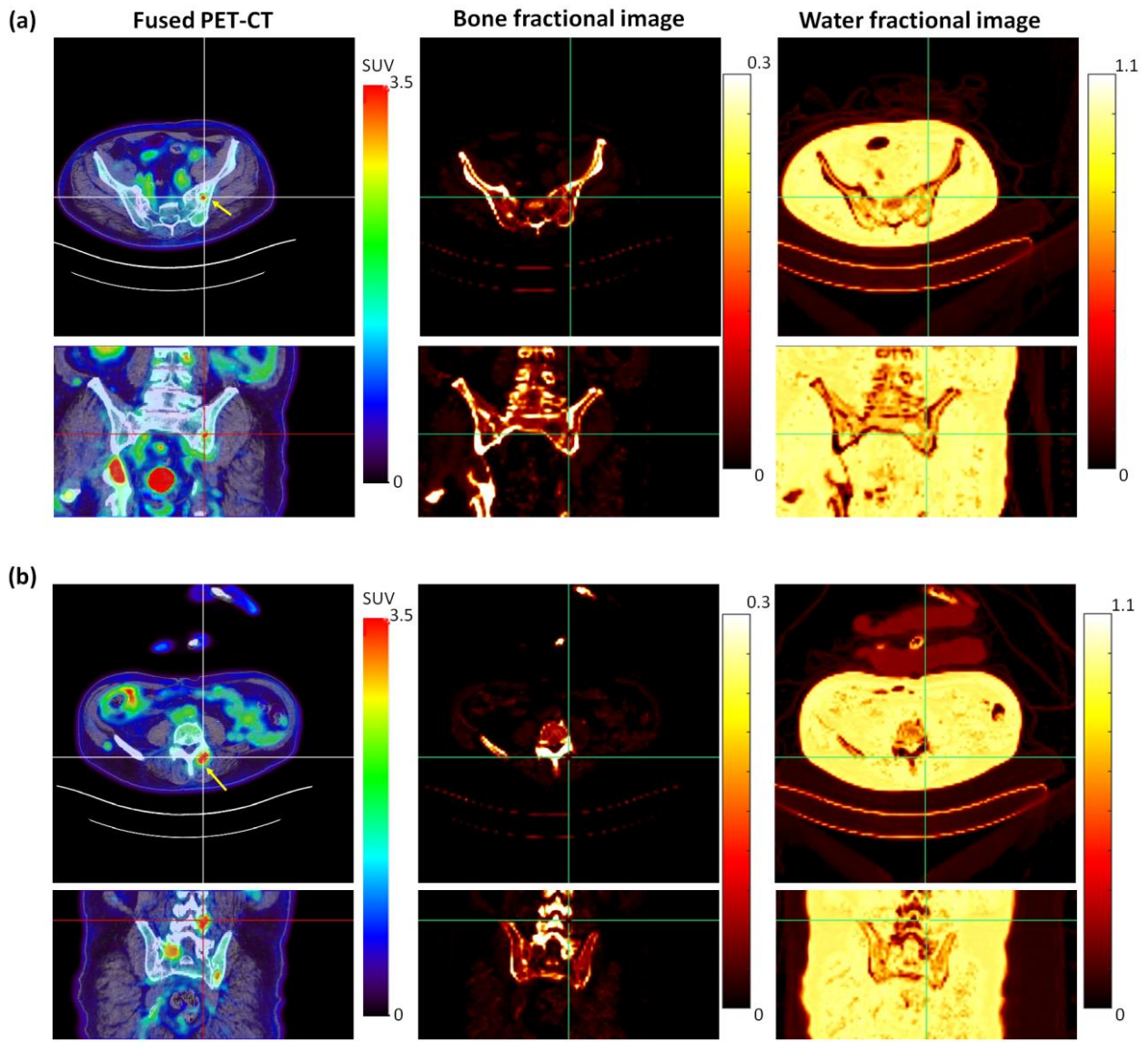

**Supplemental Fig. 8:** Transverse and coronal slices of fused PET-CT image (left column), bone fractional image (middle column) and water fractional image (right column) for (a) the second bone lesion and (b) the third bone lesion. The arrows indicate the location of the bone lesions.